\documentclass[aps,prb,preprint,groupedaddress]{revtex4-1}

\usepackage{graphicx}

\begin{document}


\title{Photonic Band Gap Effects in Two-dimensional Polycrystalline and Amorphous Structures}


\author{Jin-Kyu Yang$^1$, Carl Schreck$^2$, Heeso Noh$^1$, Seng-Fatt Liew$^1$, Mikhael I. Guy$^3$, Corey S. O'Hern$^{4,2}$, Hui Cao$^{1,2}$}


\affiliation{$^1$ Department of Applied Physics, Yale University, New Haven, CT 06520-8482}

\affiliation{$^2$ Department of Physics, Yale University, New Haven, CT 06520-8482}

\affiliation{$^3$ Science \& Research Software Core, Yale University, New Haven, CT 06520-8286}

\affiliation{$^4$ Department of Mechanical Engineering and Materials Science, Yale University, New Haven, CT 06520-8286}


\date{\today}

\begin{abstract}
We study numerically the density of optical states (DOS) in
two-dimensional photonic structures with short-range positional
order, and observe a clear transition from polycrystalline to
amorphous photonic systems. In polycrystals, photonic band gaps (PBGs) are formed within individual domains, 
which leads to a depletion of the DOS similar to that in
periodic structures. In amorphous photonic media, the domain sizes are too
small to form PBGs, thus the depletion of the
DOS is weakened significantly. The critical domain size that separates the polycrystalline and
amorphous regimes is determined by the attenuation
length of Bragg scattering, which depends not only on the degree of
 positional order but also the refractive index contrast of
the photonic material.  Even with relatively low refractive
index contrast, we find that modest short-range positional
order in photonic structures enhances light confinement via
collective scattering and interference.
\end{abstract}

\pacs{42.70.Qs, 42.25.Fx, 61.43.-j, 78.20.Bh}

\maketitle


\section{Introduction}
\label{intro}

Over the past two decades photonic crystals (PhCs) have been studied
intensely because of their ability to control light propagation and
emission \cite{joa,sou,nod}.  A unique feature of PhCs is that
they possess photonic band gaps (PBGs), within which optical modes are
absent and light propagation is prohibited \cite{yab87,joh87}.  
A PBG is an optical analog of an electronic band gap in crystals.
Typically, a PBG is formed via Bragg scattering of light by a periodic
lattice. An alternative mechanism for PBG formation is evanescent
coupling of Mie resonances of individual scatterers made of high-index
materials, which can be explained using the tight-binding model 
frequently applied to electronic band gaps in semiconductors.
Lattice periodicity or long-range order is not required, and thus many
amorphous semiconductors display large electronic band gaps.
PBGs also exist in amorphous photonic structures that consist of strong
Mie scatterers such as dielectric rods or spheres
\cite{jin01,bal99,roc06,wan06,roc09}.  These structures are
termed photonic glasses \cite{gar07}, in analogy to glassy silica,
which has an electronic band gap spanning the entire visible frequency
range. Even without Mie resonance, Bragg scattering of propagating
waves by local domains can produce PBGs in structures with only
short-range positional order.  For example, complete PBGs exist
in photonic amorphous diamond structures---three-dimensional (3D)
continuous random networks with diamond-like tetrahedral-bonding
between particles \cite{eda08}. Recently hyper-uniform disordered
materials with short-range geometric order and uniform local topology
have been shown to posses large PBGs \cite{flo09}. Unique
optical features of amorphous media have also been investigated
experimentally \cite{rec10}. Despite these studies, little is
known about the transition from PhCs to amorphous optical materials,
{\it e.g.}, how does the density of optical states (DOS) evolve as the
structural properties of the material change from ordered to
amorphous? Is there a critical size of ordered domains in
polycrystalline materials below which the system becomes optically
amorphous?  Answering the above questions will 
provide physical insight into PBG formation in structures lacking
long-range order.

In fact, nature utilizes both crystalline and amorphous photonic
structures for color generation \cite{vuk03,kin05,pru06}. 
Periodic structures are intrinsically anisotropic, thus the colors
they produce are iridescent (i.e., change with viewing
angle). In photonic polycrystals, the cumulative effect of a large number of
randomly orientated crystallites makes the color non-iridescent
\cite{kri06}. Photonic amorphous media can also produce vivid
non-iridescent colors via short-range structural order
\cite{duf09}. Although the refractive index contrast is usually too
low to form PBGs in most biological systems, the interference of
scattered light selects the color whose wavelength corresponds to the
structural correlation length \cite{noh10}. Therefore, short-range
 positional order can significantly modify photonic properties
\cite{roj04,der06,reu07}, leading to unique applications \cite{tak09}.

This paper presents numerical studies of the density of optical
states (DOS) as a function of positional order in polycrystalline and
amorphous photonic materials. To avoid Mie resonances, we consider
two-dimensional (2D) arrays of air cylinders in dielectric media,
where the PBGs are formed for transverse-electric (TE) polarized
light via Bragg scattering.  We monitor changes in the spectral region of reduced DOS as a function of the structural correlation length.  The depth of the dip in the DOS remains nearly identical to that for periodic structures as long as the ordered domains are sufficiently large. Once
the average domain size decreases below a critical value, the depth of the dip tends to zero quickly.  In contrast, the spectral width of the dip first
decreases as the ordered domains shrink, but then increases when the
domain size falls below a threshold value.  We find that the
dependence of the depth and width of the reduced DOS region on domain size $\xi$ agrees quantiatively
for different refractive index contrasts $n$ after normalizing $\xi$
by the attenuation length of Bragg scattering in a periodic
structure.  This allows us to identify the polycrystalline and
amorphous optical regimes from the ratio of the domain size
to the Bragg length. For amorphous media with low refractive index
contrast, there is little reduction in the DOS, yet scattering
is enhanced by short-range order, which results in stronger
confinement of light and higher-quality ($Q$) resonances.

The paper is organized as follows. Section II describes the methods used to generate two-dimensional polycrystalline and amorphous photonic structures, as well as a detailed characterization of the degree of spatial order. Calculation and analysis of the density of optical states in these structures are presented in Section III. Section IV demonstrates the enhanced scattering and mode confinement in systems with short-range order. Finally we conclude in Section V.

\section{Structure Generation and Characterization}
\label{gen}

\begin{figure}[htbp]
\includegraphics[width=12cm]{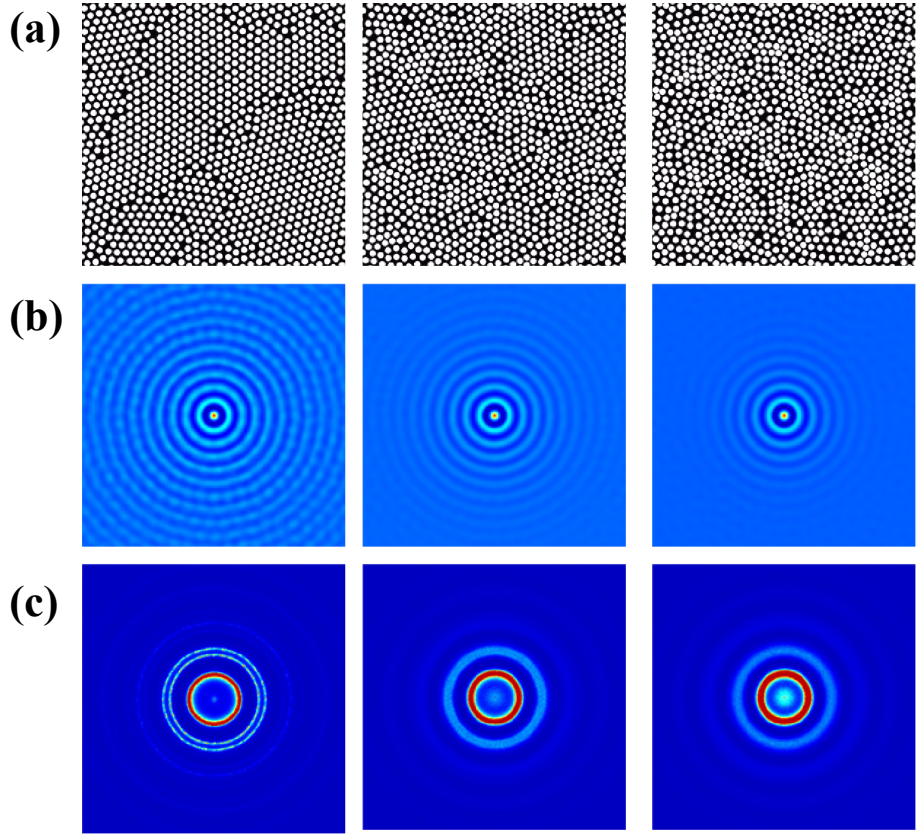}
\caption{(a) Typical configurations of two-dimensional arrays of air
cylinders (white) in a dielectric host (black), generated using
Protocol $1$ described in Sec.~\ref{gen} with polydispersity $p=0.1$
(left), $0.3$ (middle), and $0.5$ (right).  (b) Contour plot of the
ensemble-averaged density spatial autocorrelation function $C(\Delta
{\vec r})$ and (c) Power spectra $|f({\vec q})|^2$ from 
Fourier transformed density for the same polydispersities in (a).}
\label{Geom}
\end{figure}

We first describe the numerical simulation methods to generate $N$
cylinders of circular shape in a square box with periodic
boundaries. For the purpose of  generating configurations with 
varying positional order, we assume
that the cylinders interact elastically via the purely repulsive
short-range linear spring potential
\begin{equation}
\label{interaction}
V(r_{ij}) = \frac{b}{2} \left( 1 -
\frac{r_{ij}}{d_{ij}} \right)^{2} \theta \left( 1 -
\frac{r_{ij}}{d_{ij}} \right),
\end{equation}
where $r_{ij}$ is the center-to-center distance between cylinders $i$
and $j$, $b$ is the characteristic energy scale of the interaction,
$\theta(x)$ is the Heaviside function, and $d_{ij} =(d_{i}+d_{j})/2$
is the average diameter of cylinders $i$ and$j$.  To vary the degree 
of positional order, the cylinders are polydisperse -- with a uniform
distribution of diameters between $d_0$ and $d_0 (1+p)$, where $p$ is
the polydispersity that ranges from $0$ to $0.5$. The average diameter
$\langle d \rangle = d_0 (1+p/2)$. 

 Initially $d_0$, or the packing fraction
\begin{equation}
\label{phi}
\phi = \left( \frac{1}{L}  \right)^2 \sum_{i=1}^N \pi \left( \frac{d_i}{2} \right)^2, 
\end{equation}
is set to a small value $\phi_0 = 0.2$, and we place $N$ cylinders
randomly within a square of side length $L$.  We then gradually 
increase the diameters of all cylinders while maintaining the relative
size distribution to create a jammed packing of
cylinders~\cite{gao}. Each increment in diameter is followed by
minimization of the total potential energy $V=\sum_{i>j}
V(r_{ij})$ of the system. The energy minimization process is similar
to moving each cylinder along the direction of the total force on it
using overdamped dynamics. When $V$ drops below a threshold value or
the difference in energy between successive minimization steps is less
than a small tolerance, the minimization process is
terminated. If $V$ is zero and gaps exist between cylinders, the
system is unjammed, and it is compressed with a further increase of
$d_0$. If $V>0$ after the energy minimization process, a large
system-spanning number of cylinders are overlapped. To eliminate
overlap, the system is decompressed, i.e., $d_0$ is uniformly
decreased for all cylinders.  The energy minimization process is
repeated after the decompression step to find the local potential
energy minimum.  If $V=0$, the system is compressed; if not, the
system is decompressed again. The increment by which the packing
fraction of the cylinders is changed at each compression or
decompression step is gradually reduced to zero. Eventually when all of
the cylinders are just touching and the net force on each cylinder is
 nearly zero, the system is considered ``jammed'', and the
process to increase the packing fraction is stopped.

For each polydispersity $p$, we generated at least $100$ static,
jammed packings of cylinders from random initial configurations. The
values of $\phi$ are typically in the range between $0.82$ and
$0.85$ with varying degrees of positional order.  After
generating jammed packings, we reduce the diameters of all cylinders
to the same value (with $\phi = 0.5$) to eliminate the size
polydispersity. Thus, in the final configurations, the
structural disorder exists only in the positions of the cylinders with
order decreasing monotonically with increasing $p$.  Fig.~\ref{Geom} (a)
shows the typical configurations of $N=1024$ cylinders generated with
$p = 0.1$ (left), $0.3$ (middle), and $0.5$ (right). For $p=0.1$, the
system contains several domains of cylinders with crystalline
order, but each possesses a different orientation.  With increasing
$p$, the domains have reduced positional order and decrease in size.

\begin{figure}[htbp]
\includegraphics[width=15cm]{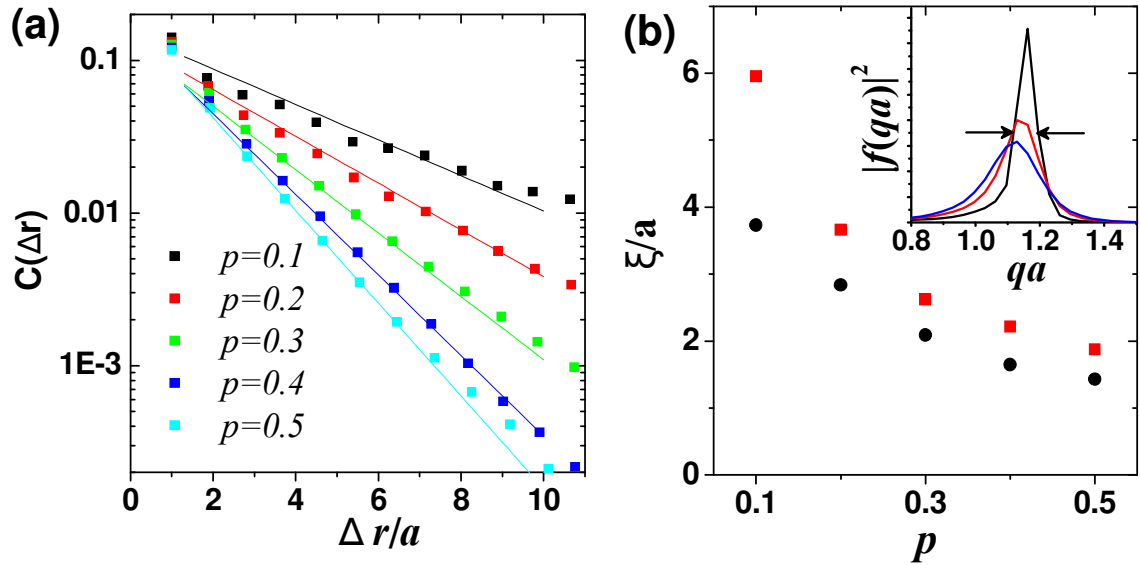}
\caption{(a) Logarithmic plot of the peak amplitudes for the ensemble- and
angle-averaged density spatial autocorrelation function $C(\Delta r)$
for $p=0.1$, $0.2$, $0.3$, $0.4$, and $0.5$. $a =L/N^{1/2}$
is the average distance between adjacent cylinders. The exponential
fits of the data (solid lines) give the decay length $\xi_r$. (b)
Inset: The first peaks of the angle- and ensemble-averaged 
Fourier transformed density $|f(qa)|^2$ for $p=0.1$ (black), $0.2$
(red), and $0.4$ (blue), whose width gives the average domain size
$\xi_q$. Main panel: $\xi_r$ (circles) and $\xi_q$ (squares) versus
$p$.}
\label{fig2}
\end{figure}

To quantify the structural order, we calculate the
ensemble-averaged spatial correlation function of the density, the
Fourier transform of the density, the radial distribution function
$g(r)$, and the local and global bond orientational order parameters.
The spatial autocorrelation function of density $\rho({\bf r}) =
L^{-2} \sum_{i=1}^N \theta({\bf r} - {\bf r}_i)$ is given by
\begin{equation}
C(\Delta \bf{r}) = \frac{ \langle \rho (\bf{r}) \rho (\bf{r} + \Delta \bf{r}) \rangle - \langle
\rho (\bf{r}) \rangle^{2}}{ \langle \rho (\bf{r}) \rangle^{2}}.
\end{equation}
$C(\Delta {\bf r})$ is averaged first over the spatial coordinates
 of the cylinders $\bf{r}$ within one configuration, and then
over at least $100$ independent configurations.  A contour plot
of $C(\Delta {\bf r})$ is displayed in Fig.~\ref{Geom} (b) as a
function of increasing $p$ (from left to right) used to generate the
configurations.  For $p=0.1$, $C(\Delta {\bf r})$ displays a large
number of concentric rings and a modulation of the amplitude within a
given ring, which indicates strong positional order.  As $p$ increases
the system becomes more disordered and isotropic, since the number of
visible concentric rings decreases and the amplitude within a given
ring becomes more uniform.  After integrating $C(\Delta {\bf r})$ over
the polar angle, we plot in Fig.~\ref{fig2} (a) the peak amplitudes of
the rings as a function of $\Delta r/a$, where $a=L/N^{1/2}$ is
the average center-to-center distance between neighboring cylinders.
The peak amplitudes decay more rapidly with $\Delta r$ at larger
$p$. The decay is approximately exponential, if we exclude the first
peak near $\Delta r = a$. The faster decay from the first peak
to the second arises from correlations induced by the
just-touching jammed cylinders. The decay length $\xi_r$ is extracted
from the exponential fit $\exp[- \Delta r / \xi_r]$ of peak amplitudes after excluding the first peak. As shown in Fig.~\ref{fig2} (b), $\xi_r$ is smaller for larger $p$, indicating the
range of spatial order becomes shorter.

We also calculated the spatial Fourier transform of the structures,
$f({\bf q}) = \int d^2 {\bf r} \exp[-i {\bf q} \cdot {\bf r}]
\rho({\bf r})$, where ${\bf q}$ is the wavevector. Figure~\ref{Geom} (c)
displays the ensemble-averaged power spectra $|f({\bf q})|^2$
for $p = 0.1$, $0.3$ and $0.5$, which consist of concentric
rings. The radial width of the rings increases with $p$, as can be
seen clearly for the first ring (with the smallest radius). The second
and third rings are distinct for $p=0.1$, which indicates the
six-fold symmetry of the cylinders within each domain. For $p=0.3$
and $0.5$, these rings become wider and merge together. We integrate
$|f({\bf q})|^2$ over all directions of ${\bf q}$ to obtain the
intensity as a function of the amplitude $q$. The inset of Fig.~\ref{fig2} (b)
displays the intensity of the first ring versus $q$ for $p = 0.1$, $0.3$
and $0.5$. The center position of the peak $q_0$ gives the dominant
spatial correlation length $s = 2 \pi / q_0$. The peak becomes broader
at larger $p$. The full width at half maximum (FWHM) of the peak
$\Delta q$ gives the average size of ordered domains $\xi_q = 2\pi/
\Delta q$. As shown in Fig.~\ref{fig2} (b), $\xi_q$ decreases with increasing
$p$, similar to $\xi_r$.

\begin{figure}[htbp]
\includegraphics[width=15cm]{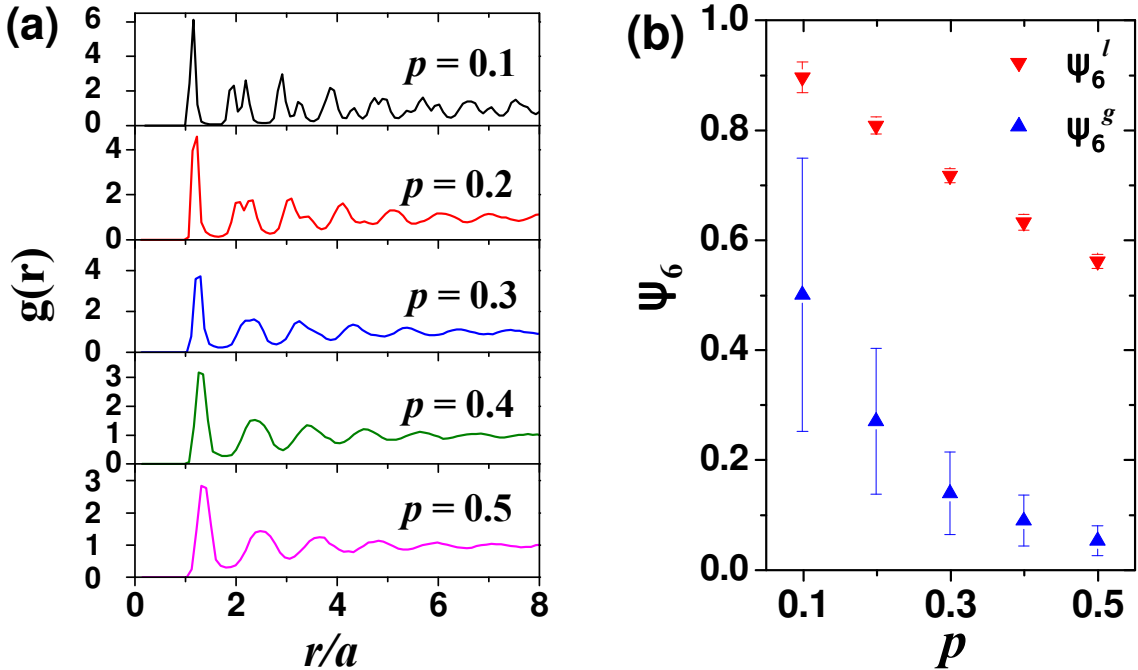}
\caption{(a) Radial distribution function $g(r)$ for $p=0.1$, $0.2$,
$0.3$, $0.4$, and $0.5$. (b) Local $\psi^l_6$ (downward triangles) and
global $\psi^g_6$ (upward triangles) bond-orientational order
parameters versus polydispersity $p$.}
\label{fig3}
\end{figure}

The radial distribution function $g(r)$, plotted in Fig.~\ref{fig3} (a) for
several values of $p$, gives the probability for a cylinder to be
located a distance $r$ from another cylinder at the origin relative to
that for an ideal gas. The strong first peak, splitting of the
second peak, and existence of peaks at large $r$ for $p=0.1$ indicate
that the structure possesses crystalline order.  With
increasing $p$, the peaks are broadened, decay faster with $r$, 
and $g(r)$ resembles that for a dense liquid~\cite{weeks}.

In addition to the translational order, we also characterized the
 orientational order of the configurations. The bond-orientational order
parameter $\psi_6$ measures the hexagonal registry of nearest
neighbors~\cite{stein}. $\psi_6$ can be calculated `locally', which
does not include phase information, or `globally', which allows phase
cancellations. Eqs.~(\ref{2dglobal}) and (\ref{2dlocal}) provide
expressions for the global and local bond-orientational order
parameters in 2D structures.
\begin{eqnarray}
\label{2dglobal}
\psi_6^{g}&=&\frac{1}{N}\left|\displaystyle\sum_{i=1}^N\frac{1}{m_i}
\displaystyle\sum_{j=1}^{m_i}e^{6\imath\theta_{ij}}\right| \\
\label{2dlocal}
\psi_6^{l}&=&\frac{1}{N} \displaystyle\sum_{i=1}^N\frac{1}{m_i}\left|
\displaystyle\sum_{j=1}^{m_i}e^{6\imath\theta_{ij}}\right|, 
\end{eqnarray}
where $\theta_{ij}$ is the polar angle of the bond connecting the
cylinder $i$ to its neighbor $j$, and $m_i$ denotes the number of
nearest neighbors of $i$. Two cylinders are deemed nearest neighbors
if their center-to-center distance $r_{ij} < r_{\rm min}$, where
$r_{\rm min}$ is the location of the minimum between the first two
peaks in $g(r)$.

As shown in Fig.~\ref{fig3} (b), both $\psi_6^{l}$ and $\psi_6^{g}$ decrease as
$p$ increases. $\psi_6^{l}$ is larger than $\psi_6^{g}$, because of
the different orientations of the ordered domains. The error bars
represent the standard deviations from $100$ configurations. For
$p=0.1$, there is a significant fluctuation of $\psi_6^{g}$, 
because some configurations have only a few distinct domains while
others contain many domains with different orientations. With
increasing $p$, the number of domains $N_d$ increases, thus the mean
and standard deviation of $\psi_6^{g}$ decrease. For $p = 0.5$,
$\psi_6^{g} \approx 0$, the structures possess only local bond
orientational order with $\psi_6^l \approx 0.55$ as found in dense
liquids~\cite{stein}.

\begin{figure}[htbp]
\includegraphics[width=10cm]{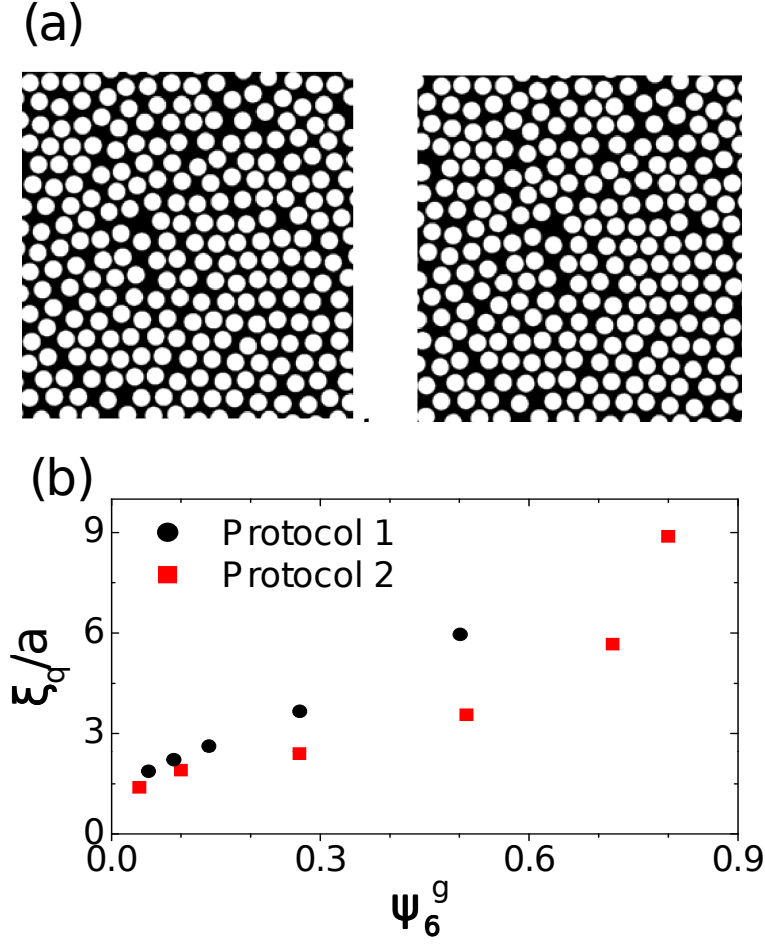}
\caption{(a) Equilibrated liquid configurations of cylinders generated
with initial packing fractions $\phi_0=0.77$ (left) and $0.73$
(right) and global bond-orientational order parameters $\psi_6^g =
0.80$ (left) and $0.27$ (right). (b) Average domain size $\xi_q$
versus $\psi_6^g$ for configurations generated using 
protocols $1$ and $2$.}
\label{fig4}
\end{figure}

To check the sensitivity of the photonic properties on the protocol
used for structure generation, we employed a second protocol that is
very different from the first. Instead of constructing jammed
packings at zero temperature, we generate equilibrated liquid
configurations of cylinders at finite temperatures. `Liquids'
exist at finite temperature with nonzero root-mean-square velocities,
 thus the net force on each cylinder, arising from the repulsive
interactions in Eq.~\ref{interaction}, does not necessarily vanish.
Employing molecular dynamics simulation methods, we simulated systems of
cylinders at constant temperature over a range of packing fractions
from $\phi_0 = 0.6$ to $0.8$, which are lower than those obtained from
the jamming protocol.  The structures at higher $\phi_0$
typically have more order, e.g., larger values of $\psi_6^{g}$. 
As in the first protocol, finally the diameters of all cylinders are reduced
to a uniform value such that $\phi=0.50$.  Figure~\ref{fig4}(a)
shows two such configurations created using different initial
packing fractions, $\phi_0 = 0.77$ (left) and $0.73$ (right). The left
(right) panel has $\psi_6^{g} = 0.80$ ($0.27$). Although the
left structure possesses stronger global order, i.e. a higher value of
$\psi_6^{g}$ than most jammed structures generated with $p = 0.1$,
 its local order parameter $\psi_6^{l}$ is smaller than most
jammed structures at $p=0.1$. This feature arises because in the
liquid state nearest cylinders do not need to remain in contact as in
jammed structures.   In contrast to the first protocol, there are
no well-defined domain boundaries in the structures generated by
the second protocol as shown for the structure on the left in
Fig.~\ref{Geom} (a).

Figure~\ref{fig4} (b) shows a plot of the average domain size $\xi_q$,
extracted from the spatial Fourier power spectra versus $\psi_6^{g}$
for the structures generated by both protocols. The variation of
$\xi_q$ with $\psi_6^{g}$ for both protocols agrees qualitatively.
 
\section{Density of Optical States}
\label{ph}

\begin{figure}[htbp]
\includegraphics[width=15cm]{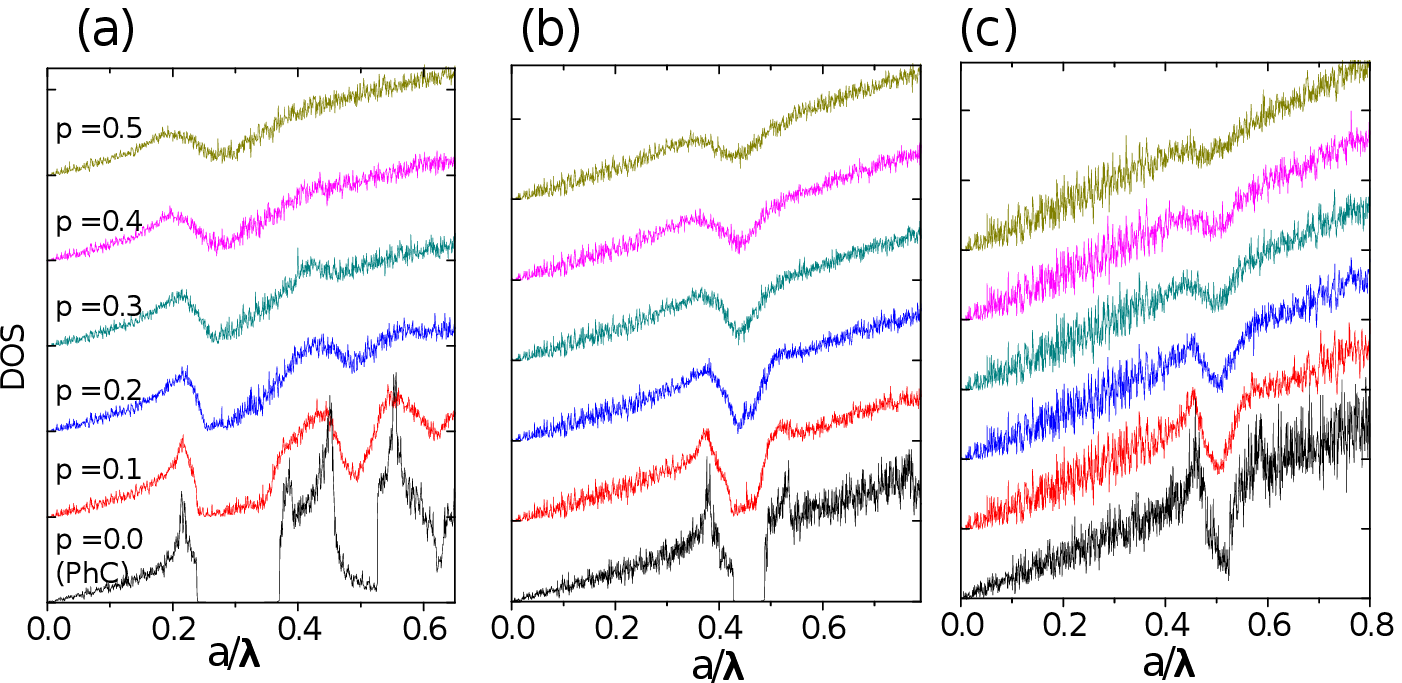}
\caption{Density of optical states (DOS) as a function of the
normalized frequency $\omega_a/2 \pi c = a/\lambda$ for the 2D
structures generated by the first protocol with $p = 0.1$, $0.2$,
$0.3$, $0.4$, and $0.5$, and a triangular lattice 
($p=0$ of identical density and diameter of air cylinders. The dielectric media, in which the air cylinders are embedded, have refractive indexes (a) $3.4$, (b) $1.8$, and (c)
$1.4$.}
\label{DOS}
\end{figure}

We calculate the DOS with transverse electric (TE)
polarization using the order-$N$ method \cite{cha95}. The magnetic
field is parallel to the axis of the air cylinders, and the
electric field exists in the 2D plane. Since the cylinders are
generated in a square with periodic boundary conditions, we can use it
as a supercell for the DOS calculation.  For the initial
conditions, we choose a superposition of Bloch waves with random
phases for the magnetic field and set the electric field to zero
\cite{lid00}. The spectral intensities, averaged over many Bloch wave
vectors and configurations, correspond to the DOS of the system under
consideration. We tested our code by reproducing the DOS for 
two-dimensional photonic structures in the literature \cite{lid00}.

\begin{figure}[htbp]
\includegraphics[width=13cm]{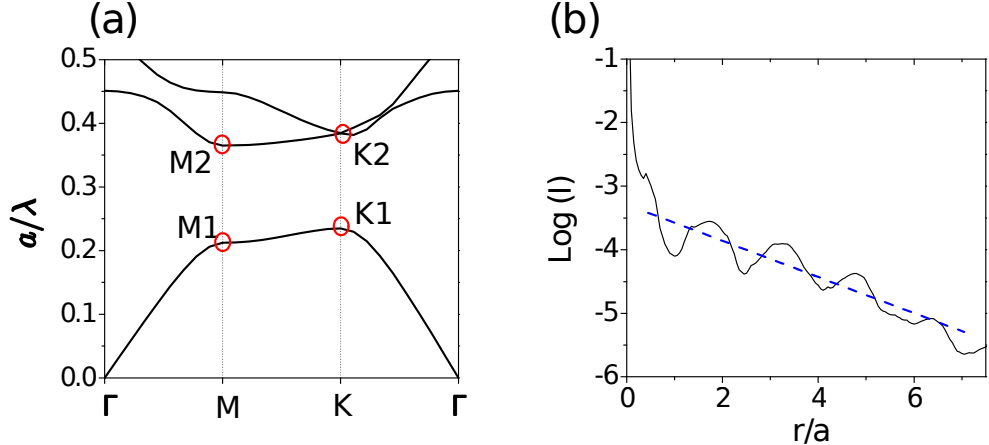}
\caption{(a) Photonic bands for a triangular array of air cylinders in a
dielectric medium of refractive index $n = 3.4$. The cylinder radius
$r = 0.37 a$, where $a$ is the lattice constant. (b) Logarithmic plot
of the angle-averaged electric field intensity $I$ versus the distance
$r$ from a dipole source oscillating at frequency $\omega_d = 0.51$ in the
 middle of a large triangular array. The dashed line is
 a fit to $\exp[-r/l_b]$, where $l_b$ is the Bragg length.}
\label{GapSW}
\end{figure}

In Fig.~\ref{DOS}, we plot the DOS as a function of the
normalized frequency $\omega a/2 \pi c = a/\lambda$ for the structures
generated by the first protocol with $p = 0.1$, $0.2$, $0.3$, $0.4$,
and $0.5$, and a triangular lattice ($p=0$) with identical density 
and diameter air cylinders. The refractive index
of the dielectric host in which the air cylinders are embedded is also
varied with $n = 3.4$, $1.8$, and $1.4$ from left to right in
Fig.~\ref{DOS}.  For $n = 3.4$ and $p = 0$, a complete depletion of
the DOS from $a/\lambda = 0.235$ to $0.365$ results from the full PBG
between the first and second bands of the triangular lattice  
[Fig.~\ref{GapSW}(a)]. With the introduction of positional
disorder, defect modes are created inside the gap, and the frequency
region of depleted DOS becomes shallower and narrower. The higher frequency
side of the gap (air band edge) is affected more than the lower
frequency side (dielectric band edge). Because the air holes are
isolated and the dielectric host is connected, the dielectric bands
below the gap are more robust to the disorder than the air bands above
the gap. For $n = 1.8$, the PBG of the periodic structure becomes
smaller, and thus the depleted region of the DOS is narrower. For
the perfect crystal with $n=1.4$, the first photonic band at the $K$
point ($K1$) has the same energy as the second band at the $M$ point
($M2$), thus the full PBG disappears.  As a result, the DOS
displays a dip, rather than a complete depletion.  As shown in
Fig.~\ref{DOS}, the addition of positional disorder causes the dip in
the DOS to become shallower and eventually disappear at large disorder.

\begin{figure}[htbp]
\includegraphics[width=10cm]{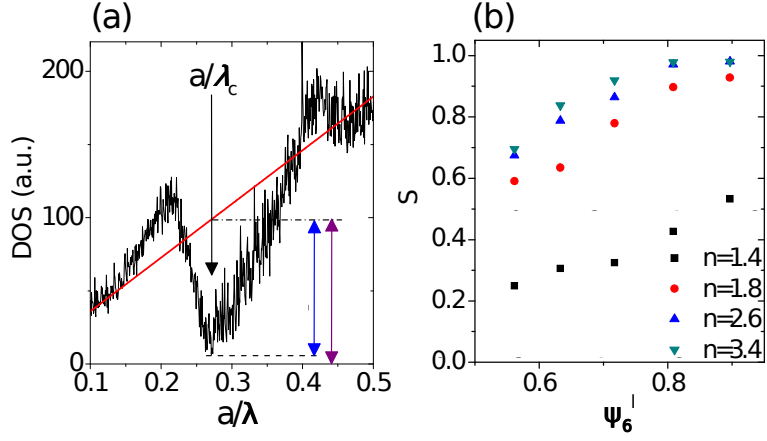}
\caption{(a) A schematic defining the normalized depth $S$,
 which is the ratio of the maximal reduction depth of the DOS
(black curve) at frequency $a/\lambda_c$ (blue segment with arrows) to
the DOS of a random structure (fitted by the red solid line) at the
same frequency (purple segment with arrows). (b) $S$ versus the local
bond-orientational order parameter $\psi_6^l$ for arrays of air
cylinders in a dielectric host of refractive index $n = 1.4$, $1.8$,
$2.6$, and $3.4$.}
\label{Liquid}
\end{figure}

To quantify the strength of the DOS depletion, we introduce the
normalized depth $S$, which is defined as the ratio of the
maximal depth of DOS reduction to the DOS of a random structure at the same
frequency [Fig.~\ref{Liquid}(a)]. The density and diameter of air cylinders as well as the
refractive index of the dielectric host in the random structure are
identical to those of the structures under investigation. The DOS of
the random structure increases almost linearly with frequency, similar
to that of a homogeneous 2D dielectric medium. A linear fit of the DOS
is shown as the red line in Fig.~\ref{Liquid}(a).
We investigated the dependence of $S$ on various order parameters,
e.g. the local bond orientational order $\psi_6^{l}$. As shown in
Fig.~\ref{Liquid}(b), $S$ increases gradually with $\psi_6^{l}$. 
However, the variation depends on the refractive index contrast $n$,
and is therefore not universal.

To obtain universal behavior for a given degree of positional order, 
we must account for the effect of refractive index contrast on the DOS. 
The refractive index contrast determines the strength of the PBG, which is reflected in the
attenuation length of Bragg diffraction, or the Bragg length
$l_b$. Roughly speaking, the Bragg length gives an order of
magnitude estimate for the minimal size of a periodic structure that
is necessary to form a PBG via Bragg scattering.  Since periodic
structures are anisotropic, $l_b$ varies with direction.  
However, since the DOS is a sum of optical modes in all directions,
the relevant Bragg length is an average over all directions.  To
obtain the value of $l_b$ in the numerical simulations, we place a continuous
dipole source of frequency $\omega_d$ in the middle of a large 
triangular array of air cylinders. We then calculate the electic field 
intensity at a distance $r$ from the source, and integrate it over the polar
angle. The Bragg length $l_b$ is extracted from the exponential decay of the angle-integrated field intensity $I$ with $r$, as shown in Fig.~\ref{GapSW} (b).

\begin{figure}[htbp]
\includegraphics[width=10cm]{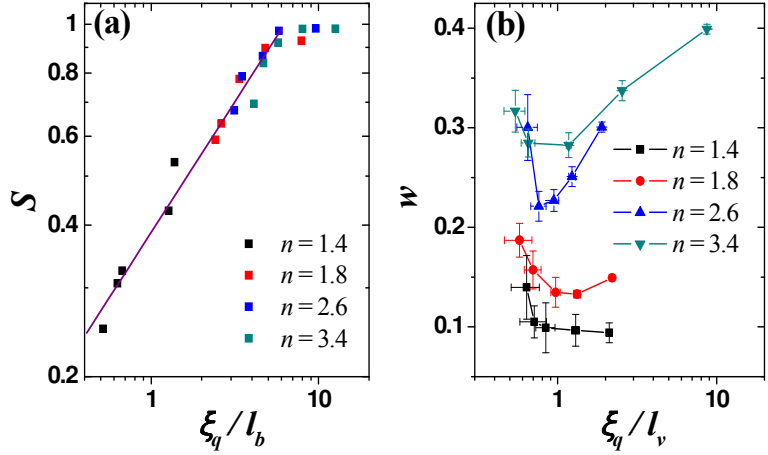}
\caption{(a) Normalized depletion depth of the DOS $S$ for arrays of
air cylinders in a dielectric host of refractive index $n$ versus
the ratio of the average domain size $\xi_q$ to the angle-averaged
Bragg length $l_b$. A linear fit (solid purple line) of the data 
on the log-log plot for $\xi_q/l_b \lesssim 5$ gives a power-law
scaling exponent of $0.52$. (b) Relative width $w$ of
the frequency region where there is a reduction in the DOS as a function of
$\xi_q/l_v$, where $l_v$ is the frequency- and angle-averaged Bragg
length. The error bars are obtained from the standard deviation
of $\xi_q$ for different configurations and fitting errors in the FWHM
of the DOS reduction zones.}
\label{MFP}
\end{figure}

In Fig.~\ref{MFP}(a), we plot the depth $S$ versus the
average size of the ordered domains normalized by the Bragg length,
$\xi_q/ l_b$. All data points for different refractive index
contrasts fall on a single curve.  When $\xi_q/l_b$ is above a
threshold value ($\sim 5$), $S$ is almost unity, which implies
that the depletion in the DOS is nearly complete as in a perfect
crystal. When $\xi_q/l_b \lesssim 5$, $S$ decreases rapidly.
The drop can be fit by a straight line on a log-log plot, which
reveals a power-law decay with an exponent $\sim 0.52$. This result
can be understood qualitatively as follows. If the domain size is
larger than the Bragg length, Bragg scattering in a single domain is
strong enough to form a PBG. The DOS in systems with large $\xi_q/l_b$ 
is nearly equal to the DOS of a perfect crystal, and these structures can be
regarded as photonic polycrystals.  In addition, an average over many
domains of different orientations makes the directional DOS isotropic. 
If the domain size is smaller than the Bragg length, individual domains 
are too small to form PBGs. In this case, the effect of Bragg scattering is reduced due to a
limited number of periodic units, and the depletion of the DOS
is weakened. This is the amorphous photonic regime, where
short-ranged order leads to a partial depletion of the
DOS. The well-defined threshold in $\xi_q/l_b$ demonstrates a
clear and sharp transition from polycrystalline to amorphous photonic
structures.

In addition to the depth of the DOS reduction, we also studied the
spectral width of the reduction region. The relative width $w$ is 
defined as the ratio of the full width at half minimum (FWHM) of the
dip in the DOS $\delta \omega$ to the frequency $\omega_0$ at the
center of the dip. Since the Bragg length varies within the spectral
region of DOS reduction, we average its value over the frequency range from $\omega_0 -
\delta \omega$ to $\omega_0 + \delta \omega$. The average domain size
is normalized by the average Bragg length $l_v$. Figure~\ref{MFP} (b)
 shows a plot of $w$ versus $\xi_q/l_v$ for several values of
$n$. Although the curves for different $n$ do not coincide, their
trends are similar. As $\xi_q/l_v$ increases, $w$ first drops and then
rises (except for $n = 1.4$). The turning point is at $\xi_q/l_v \sim 1$.
To understand this behavior, we first examine the DOS for
periodic systems. At $n = 3.4$, the DOS is enhanced at the photonic
band edges due to the slow group velocity [Fig.~\ref{DOS} (a)]. When
positional disorder is introduced to the structure, the DOS peak at
the air band edge is quickly lowered and the higher frequency part of
the PBG is filled by defect modes. In contrast, the peak at the
dielectric band edge decreases more slowly, because the dielectric
bands are more robust against disorder as mentioned earlier. The gap
width is reduced, until the DOS peak at the dielectric band edge
diminishes at a certain degree of disorder. Then the DOS below the
dielectric band edge starts decreasing with further increases in
disorder. The DOS reduction region becomes wider. As $n$ decreases,
the strength of DOS reduction by PBGs is weakened, and $\delta \omega/
\omega_0$ is lowered. At $n=1.4$, $w$ no longer rises beyond $\xi_q/l_v
\sim 1$; instead it tends to a plateau.

\begin{figure}[htbp]
\includegraphics[width=10cm]{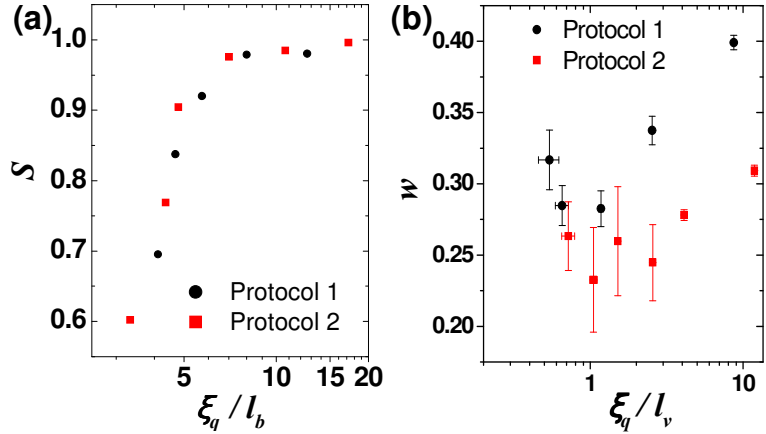}
\caption{
Normalized depth $S$ (a) and width $w$ (b) of the DOS reduction region for the structures generated by protocols $1$ (circles) and $2$ (squares). The refractive index of the dielectric host, within which air cylinders, is $n=3.4$. The horizontal axes are indentical to those in Fig. 8 (a) and (b).  The error bars in (b) are obtained using the same method in Fig.~\ref{MFP}(b).
}
\label{fig9}
\end{figure}

To check the robustness of our results on the method for
generating configurations, we also calculated the DOS for
structures produced by the second protocol. Figure~\ref{fig9}
compares the normalized depth $S$ and relative width $w$ of the
DOS reduction regions for the structures
generated by the two protocols at $n=3.4$.  All data points for $S$
versus $\xi_q/l_b$ fall on the same curve in Fig.~\ref{fig9} (a). The 
data for $w$ versus $\xi_q/l_v$ for the two protocols display a similar 
trend as shown in Fig.~\ref{fig9} (b); however, the turning point
is slightly shifted, and the values of $w$ beyond the turning point
are smaller for the structures produced by the second protocol. 
A more quantitative comparison will be carried out in the future.

\section{Enhanced scattering and mode confinement by short-range order}
\label{enhanced}

\begin{figure}[htbp]
\includegraphics[width=10cm]{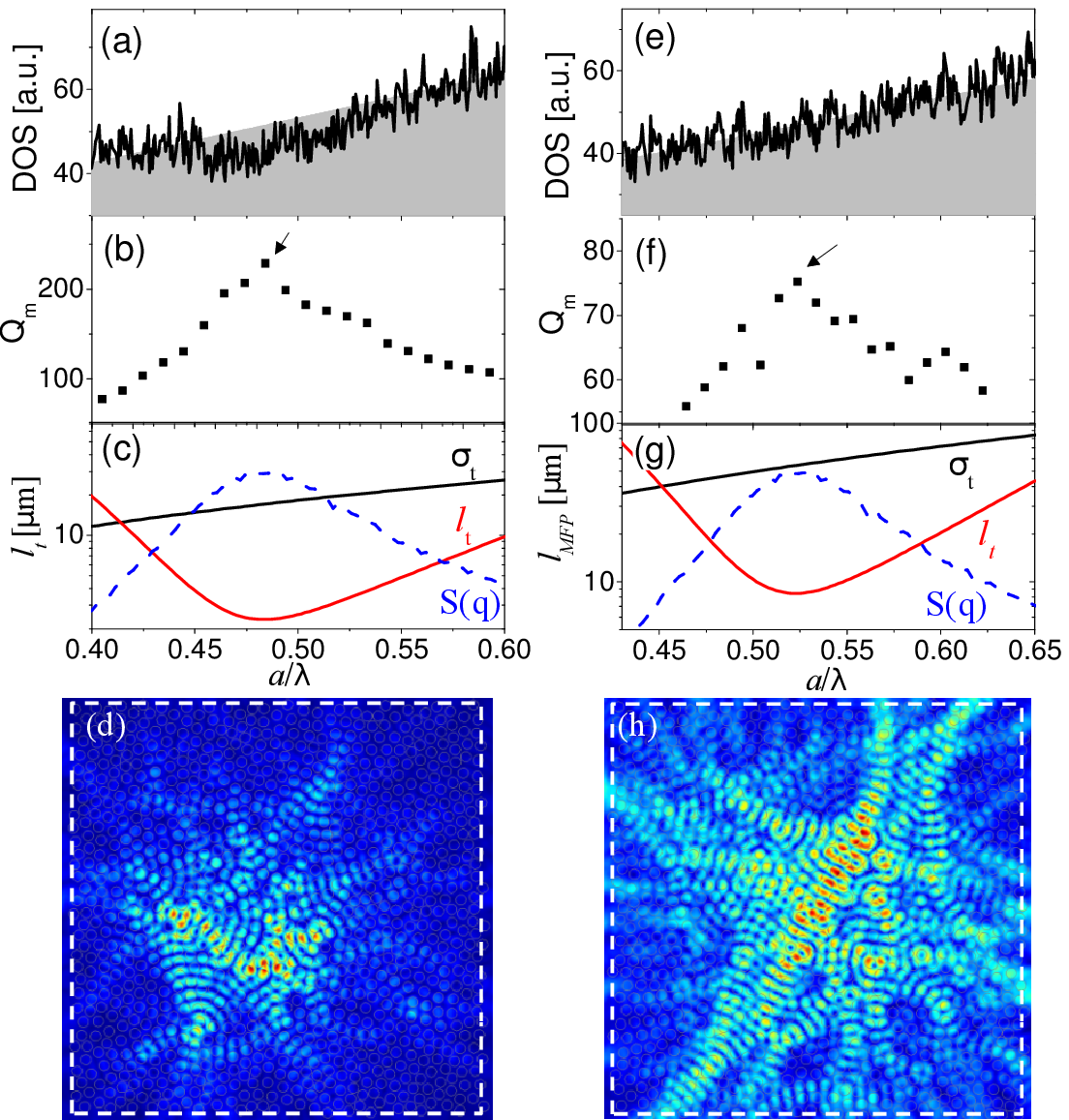}
\caption{The DOS (a,e) and maximal quality factors $Q_m$ of resonant
modes (b,f) for the amorphous photonic structures with low refractive
index contrast $n$ = 1.4 for (a-d) and 1.2 for (e-h). Grey backgrounds
in (a,e) represent the DOS for 2D homogeneous media.  (d,h): Spatial
distribution of electric field intensities for the modes of maximal
$Q_m$ [marked by arrows in (b,f)]. (c,g):  Transport mean free path
$l_t$ (red solid line), total scattering cross sections of a single
scatterer $\sigma_t$ (black dotted line), and the structure factor
$S(q)$ at $q = 2k$ (blue dashed line), where $k$ is the wavevector 
of light.}
\label{fig10}
\end{figure}

In nature, the refractive index contrast is typically low,
nevertheless photonic amorphous structures are used to manipulate light
scattering and color generation. In this section, we investigate the
effects of short-range order on light scattering and mode confinement
in amorphous structures with low index contrast. We consider the
structures generated by the first protocol with $p = 0.5$.  When
we set $n=1.4$, the DOS possesses an extremely shallow dip as shown in
Fig.~\ref{fig10}(a). For $n=1.2$, the DOS in Fig.~\ref{fig10}(e) 
is nearly featureless. We calculate the resonant modes in 
these structures using the finite element method. Instead of
periodic boundary conditions, the structures have finite size and open boundaries. Each
structure contains $1024$ air cylinders in a dielectric medium. The open boundaries
are terminated by perfectly matched layers that
absorb all outgoing waves. Because of light
leakage from the finite-sized structure, the resonant modes have
finite lifetimes. We calculate the complex frequencies of all
resonances $\omega_r + i \omega_i$. The amplitude of $\omega_i$ is
inversely proportional to the lifetime. The quality factor is defined
as $Q = \omega_r / 2 \left| \omega_i \right|$. We obtain the maximal
quality factors $Q_m$ of modes within small frequency intervals, and
plot them in Fig.~\ref{fig10}(b,f). Although the dip
in the DOS is barely visible at $n=1.4$, $Q_m$ is enhanced by a factor
of three at a frequency near the center of the dip.  Further,
even though there is essentially no dip in the DOS for $n=1.2$, $Q_m$ displays a
peak. Figure~\ref{fig10} (d,h) shows the spatial
distributions of electric field intensities $|E(x,y)|^2$ for the modes
with maximal $Q_m$ (marked by the arrows in
Fig.~\ref{fig10}(b,f). It is evident that the mode of maximal
$Q_m$ at $n=1.4$ is localized within the structure. For $n=1.2$ the
mode is more delocalized, but the field intensity near the boundary (marked by white dashed line) is
still weaker than that in the interior. To determine the degree of 
localization, we calculate the inverse participation ratio for these two modes,
\begin{equation}
s \equiv {{1} \over {L^2}} {{\left( \int |E(x,y)|^2 dx dy \right) ^2}
\over {\int |E(x,y)|^4 dx dy}},
\end{equation}
 where a mode uniformly distributed over the sample gives $s=1$.
We find that the mode in Fig.~\ref{fig10}(d) has $s=0.18$ and is thus highly
localized, while the one in Fig.~\ref{fig10}(f) has $s=0.44$ and is only
partially localized.

To illustrate the physical mechanism that leads to mode confinement,
we calculate the transport mean free path
\begin{equation}
{{1} \over{l_t}} = {{\pi} \over {k^6}} \int_0^{2k} \rho F(q) S(q) q^3 dq,
\end{equation}
where $k$ is the wavevector of light, $\rho$ is the number density of air
cylinders, $S(q)$ is the structure factor, $F(q)$ is the form factor,
and $q$ is the spatial frequency. $F(q)$ is given by the differential
scattering cross section of a single air cylinder in the dielectric
medium. The structure factor is given by
\begin{equation}
S({\bf q}) \equiv {{1} \over {N}} \sum_{i,j=1}^N e^{i {\bf q} \cdot
({\bf r}_i - {\bf r}_j)},
\end{equation}
where ${\bf r}_{i}$ denotes the center position of the $ith$
cylinder. Since the structures are isotropic, $S({\bf q})$ is
invariant with the direction of ${\bf q}$ and is only a function of
the magnitude $q$. In Fig.~\ref{fig10}(c,g), we show that
$l_t$ displays a significant drop at a frequency that coincides with
the peak in $Q_m$. This indicates that the enhancement of scattering
strength improves mode confinement. In Fig.~\ref{fig10}(c,g) 
we also plot the total scattering cross section $\sigma_t$ of a
single air cylinder, which increases monotonically with frequency and
does not exhibit any resonant behavior within the frequency range
studied.  This behavior suggests that the dip in $l_t$ is not
caused by Mie resonance of individual scatterers. Instead, we
contend that the short-range order enhances Bragg backscattering at
certain wavelengths and shortens $l_t$. To prove this, we also plot $S(q)$ for the
backscattering $q=2 k$ in Fig.~\ref{fig10} (c) and (g).  $S(q)$ is
peaked at the dip of $l_t$, which confirms that collective
backscattering from local domains of ordered cylinders causes 
a dramatic decrease in $l_t$. Therefore, the
spatial confinement of resonant modes is enhanced by short-range order
through constructive interference of scattered light that occurs at
specific frequencies.

\section{Conclusion}
\label{conclusion}

We generated polycrystalline and amorphous photonic structures
using two protocols: one that produced jammed packings of cylinders
and another that sampled equilibrated liquid states of cylinders.
 The degree of positional order was fully characterized by spatial
correlation functions, Fourier power spectra, radial distribution
functions, and bond-orientational order parameters. We were
able to gradually decrease the average domain size and trace the
transition from polycrystalline to amorphous media. The properties of
the DOS were calculated for air cylinders embedded in a dielectric
host with a refractive index contrast that varied from $n=1.4$ to
$3.4$. For $\xi_q/l_b \gtrsim 5$, where $\xi_q$ is the average
domain size and $l_b$ is the angle-integrated Bragg length, the
maximal depth of the DOS depletion is nearly unchanged from that for
the crystalline structure. This is the polycrystalline regime, where
individual domains are large enough to form PBGs. When $\xi_q/l_b
\lesssim 5$, Bragg scattering from single domains is too weak to form
PBGs, and the gap in the DOS diminishes. The spectral region of the DOS
reduction first narrows with increasing disorder, then widens when the
average domain size becomes less than the angle- and frequency-averaged Bragg
length. These results can be explained by the increase of the DOS inside
the gap due to defect modes and the decrease of the DOS outside the
gap once the band edge modes with slow group velocities are destroyed
by disorder. The behavior of the DOS is similar for  
structures generated by two qualitatively different protocols. 

We also investigated the enhancement of light scattering and mode
confinement in photonic amorphous structures with low-refractive index
contrast. Even though the PBG effect is barely visible in the
DOS, the transport mean free path displays a dramatic reduction and
the quality factors of resonances reach a maximum. The
short-range structural order enhances collective scattering of light
and improves mode confinement. Therefore, in addition to the
form factor or Mie resonances of individual scatterers, tailoring the
structure factor is an efficient way of manipulating light scattering
and confinement in photonic materials.

The authors thank Profs. Eric R. Dufresne and Richard O. Prum for
stimulating discussions. This work is funded by NSF Grant
Nos. DMR-0808937 (HC) and DMS-0835742 (CO, CS), and a seed grant from the
Yale MRSEC (DMR-0520495). J.-K. Yang acknowledges the support of the
National Research Foundation of Korea Grant funded by the Korean
Government [NRF-2009-352-C00039].  This work also benefited from the
facilities and staff of the Yale University Faculty of Arts and
Sciences High Performance Computing Center and NSF grant
No. CNS-0821132 that partially funded acquisition of the computational
facilities.

\end{document}